\documentclass[lettersize,journal]{IEEEtran}
\usepackage{amsmath,amsfonts}
\usepackage[linesnumbered,ruled,vlined]{algorithm2e}
\SetKwComment{Comment}{/*}{}
\usepackage{array}
\usepackage[caption=false,font=normalsize,labelfont=sf,textfont=sf]{subfig}
\usepackage{textcomp}
\usepackage{stfloats}
\usepackage{url}
\usepackage{xurl}
\usepackage{hyperref}
\usepackage{verbatim}
\usepackage{graphicx}
\usepackage{cite}
\def\BibTeX{{\rm B\kern-.05em{\sc i\kern-.025em b}\kern-.08em
    T\kern-.1667em\lower.7ex\hbox{E}\kern-.125emX}}
\usepackage{balance}
\usepackage{multirow}
\usepackage{booktabs}
\usepackage{threeparttable}
\usepackage{arydshln}

\usepackage{makecell}
\usepackage[english]{babel}
\usepackage{amsthm}

{ 
\theoremstyle{definition}

}

\usepackage{xcolor}

\begin{document}

\title{B2G-Agent: An LLM-Based Multi-Agent Tool for End-to-End Building-Grid Interactions}

\author{Xueyuan Cui, \IEEEmembership{Member, IEEE,} Qian Zhang, \IEEEmembership{Graduate Student Member, IEEE,}\\Xiaochen Liu, \IEEEmembership{Member, IEEE,} and Yi Wang, \IEEEmembership{Senior Member, IEEE}
}

\markboth{Submitted to IEEE Trans. Smart Grid}%
{Shell \MakeLowercase{\textit{et al.}}: A Sample Article Using IEEEtran.cls for IEEE Journals}


\maketitle
\begin{abstract}
Buildings are major electricity consumers and can interact with distribution network dispatch through flexible power scheduling. These interactions require distribution network operators (DNOs) and building energy managers (BEMs) to reconcile their operational constraints and perform complex cross-field computations. The lack of cross-domain expertise and unified computational workflows could lead to inefficient or physically infeasible interactions. To address this bottleneck, we develop an end-to-end large language model (LLM)-based multi-agent tool, \textit{B2G-Agent}, that bridges DNOs and BEMs. The tool deploys three collaborative agents: a Translator for bidirectional cross-domain semantic mapping via hybrid retrieval-augmented generation (RAG); an Actor for automated configuration and execution of joint simulation and optimization through directed acyclic graphs; and a Reviewer for memory-augmented reflective failure diagnosis. This workflow abstracts away cross-field modeling complexities and enables both sides to focus on high-level operational decisions. Comprehensive evaluations across 200 synthesized tasks show that B2G-Agent improves the interaction quality score from 66.71 to 85.57 and increases the task success rate from 38.0\% to 69.0\% over the baseline while satisfying operational constraints. The open-source tool and interface are available at \url{https://github.com/hkuedl/B2G-Agent}.
\end{abstract}


\section{Introduction}\label{sec_Intro}
As electrification increases, buildings account for more than half of global electricity consumption~\cite{newIEABuildings2026}. Along with their large electricity consumption, buildings are also becoming important flexible resources for power systems on the demand side~\cite{newliu2022photovoltaics}. Building managers can adjust the schedules of their electrical devices to support various distribution-network dispatch tasks through demand response, such as peak shaving, voltage management, and congestion management~\cite{newIEABuildingFlexibility2024}.

Utilizing building flexibility for distribution network dispatch requires interactions between the distribution network operator (DNO) and building energy managers (BEMs). In this work, interaction refers to information exchange between the two sides through communication, without considering the physical implementation of power scheduling or network dispatch using specific control infrastructure. Taking a DNO-issued dispatch task as an example of information exchange, the DNO specifies the required power adjustment, target locations, and time period; each BEM then assesses its building's operating conditions, determines a feasible adjustment, and reschedules energy resources to deliver the committed flexibility; finally, the DNO evaluates whether the combined response meets the dispatch objective and determines each participant's payment based on delivery and settlement rules~\cite{newSSENFlexTerms2025}. A complete interaction thus involves multiple intermediate steps requiring joint problem formulation and computation that account for the DNO's flexibility needs, operating states, and network constraints, as well as the schedules and occupants' living requirements of buildings by BEMs.

Building-grid interactions are predominantly conducted by humans through two main arrangements. The first involves individual BEMs communicating their flexibility responses directly to the DNO through a flexibility market platform~\cite{newSSENFlexTerms2025}. Specifically, each BEM submits flexibility based on the DNO's requirements and follows dispatch instructions. This arrangement is primarily applicable to large commercial and office buildings. The second involves an aggregator that coordinates buildings and submits their combined flexibility response to the DNO~\cite{newSSENFlexTerms2025}. Typical implementations include remote control of residential devices and coordinated scheduling through energy management systems~\cite{newTurnerHeatPumpFlex2025,newYuResidentialCoordination2024}. This arrangement is mainly applied to large numbers of dispersed residential buildings, enabling their collective participation in network dispatch. Although these two arrangements facilitate building-grid interactions across different building types, both rely on specialist humans to execute intermediate steps, which are technically complex and labor-intensive. Moreover, as these steps involve both distribution network operation and building energy management, their execution requires coordination between the two sides of humans and a mutual understanding of their domain-specific knowledge.

In recent years, large language model (LLM)-based agents have been developed to assist humans in handling complex tasks across the building and power system domains. In power systems, \cite{pf1,pf2,pf3} performed power flow analysis by training a reasoning model, generating scripts for customized distribution optimal power flow, and integrating tool execution with verification and user feedback, respectively. For simulation tasks, \cite{sim1,sim2} developed multi-agent frameworks that leverage execution feedback and persistent function memory to repair simulation code, respectively. \cite{gen1} employed a tool-augmented LLM to construct power grid models. For network dispatch, \cite{dis1,dis2,dis3} formulated optimization problems and generated solution code through hierarchical planning and action. \cite{cps1} employed LLM agents to generate attack and defense strategies for cyber-physical power systems. In building energy applications, \cite{em2} coordinated hierarchical agents and analysis tools for building energy data analysis; \cite{em3} generated EnergyPlus models from building descriptions and debugged them using simulation feedback; and \cite{ema1} integrated EnergyPlus with memory-enabled agents to balance energy costs and occupant comfort in energy management. These agents primarily automate problem formulation and execution within a specific domain, while their functions focus on interpreting the natural-language inputs of a single human. Beyond these studies, several emerging approaches have explored both power and building domains. OptAgent~\cite{ema2} coordinated specialized agents in a physics-informed simulation environment to assess building operational flexibility, while EnergyBridge~\cite{fe} incorporated resident authorization to evaluate the feasibility and delivery of flexibility. AutoB2G~\cite{ema3} constructed and executed co-simulation workflows connecting building control and distribution network models. These studies enable cross-domain computation and account for building-side participant responses. However, their interactions primarily focus on BEMs, while the power network is mainly represented through predefined models, operating constraints, or simulation feedback.

In summary, existing LLM-based agents exhibit two functional limitations. First, they primarily interpret and execute the requirements and preferences of an individual human, rather than supporting the coordinated execution of decisions involving multiple humans from different fields. Second, they focus on problem formulation and computation within a single domain, or incorporate another domain through predefined constraints or external signals, without enabling joint decision-making across domains. These functional limitations hinder their application to end-to-end building-grid interactions. Here, end-to-end interaction means that human operators only need to provide task requirements and feedback, while agents autonomously complete the intermediate steps, such as problem formulation, simulation, optimization, and verification, thereby improving interaction efficiency.

\textit{How can LLM agents facilitate end-to-end decision-making between BEMs and DNOs under cross-domain operational constraints?} To address this, we propose an LLM-based tool that integrates a multi-agent workflow applied to different interaction tasks. Through coordinated functions, these agents translate requests from one side into instructions for the other and automatically configure and execute the required computational tasks, allowing both sides to focus on their own responsibilities while the agents handle complex intermediate steps. The main contributions are summarized as follows:

\begin{enumerate}
\item We develop B2G-Agent as an LLM-based multi-agent tool for end-to-end interactions between DNOs and BEMs. This tool implements critical steps in interactions automatically, including request interpretation and execution, power scheduling, and impact assessment, through a structured message protocol. It abstracts away cross-field modeling and computing complexities and enables both DNOs and BEMs to focus on high-level operational decisions.

\item We design a coordinated workflow with three specialized agents named Translator, Actor, and Reviewer to enable end-to-end interactions. The Translator employs role-conditioned hybrid retrieval-augmented generation (RAG) to generate semantically aligned instructions; the Actor applies backward task decomposition to orchestrate cross-field simulation and optimization; the Reviewer combines semantic and evidence validations to identify and correct failures, with adaptive updates of reflection memory.

\item We release B2G-Agent as an open-source software tool on GitHub~\cite{b2g-agent}, including a graphical user interface, knowledge bases, and execution records. We construct a reproducible study comprising 200 synthesized interaction tasks across seven building–grid interaction categories. Comparisons with a single-agent baseline and agent-specific ablations demonstrate the effectiveness of the tool in interaction quality and agent functions.



\end{enumerate}

The remainder of this paper is organized as follows. Section II formulates the building-grid interaction problem. Section III presents the proposed multi-agent tool. Section IV presents the case studies. Section V concludes the work.

\section{Problem Statement}\label{sec_Pro}

\begin{figure}[t]
    \centering
    \includegraphics[width=1.0\columnwidth]{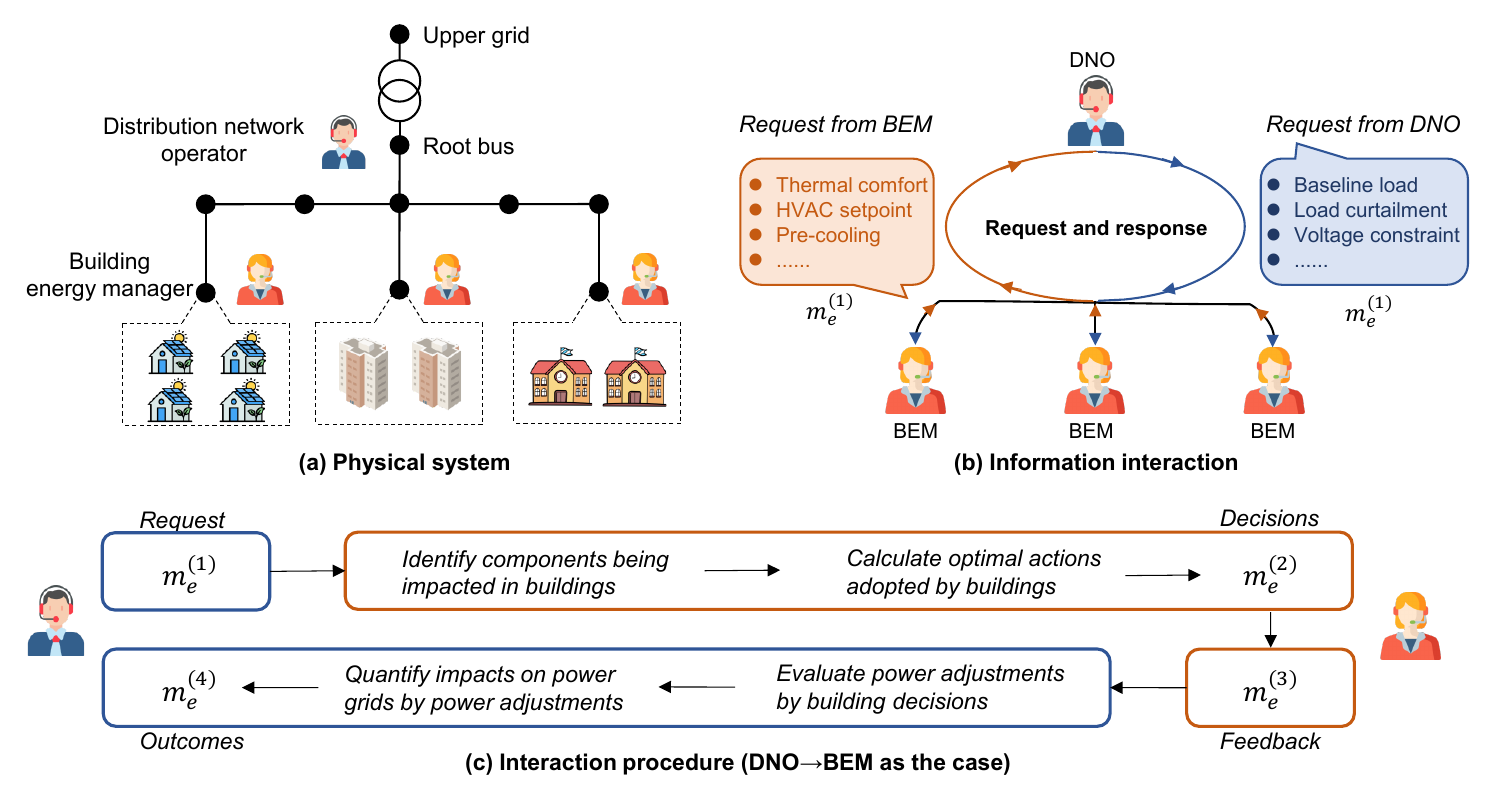}
    \caption{Physical configuration, bidirectional information exchange, and decision-making workflow for building–grid interactions.}
    \label{problem}
    \vspace{-1.0em}
\end{figure}

In Fig.~\ref{problem}, we present the research problem of building-grid interaction, including its physical configuration, cases of information interactions, and the detailed interaction procedure. In particular, Fig.~\ref{problem}a illustrates the physical settings with a DNO and multiple BEMs. Each BEM manages a large building or a building cluster connected to a single network bus. The DNO is responsible for reliable network operations and uses mechanisms such as demand response to encourage BEMs to adjust their electricity consumption. Each BEM manages flexible energy use in buildings and supports grid operation while maintaining internal occupant demands.

A complete bidirectional interaction is denoted by $e$ and consists of four sequential messages. Either DNO or BEM can act as the sender and submit an initial request $m_e^{(1)}$ for their own purpose. As illustrated in Fig.~\ref{problem}b, a BEM may report a building baseline, available flexibility, or a bid for energy arbitrage; the DNO may request peak shaving, congestion management, voltage management, or renewable energy integration. Taking the DNO as the sender as an example, Fig.~\ref{problem}c illustrates the subsequent processing of interactions. The BEM identifies the operational objectives and other key information in $m_e^{(1)}$ and produces an instruction $m_e^{(2)}$ based on the buildings' operational states. The BEM then considers its own operational conditions and responds to $m_e^{(2)}$ with $m_e^{(3)}$ by accepting, modifying, rejecting, or requesting clarification. Reverse cross-domain processing converts $m_e^{(3)}$ into $m_e^{(4)}$, which represents the impacts on power grids from each BEM's actual decisions.

The above process reveals two major issues in the building-grid interaction:
\begin{itemize}
    \item \textbf{Cross-domain semantic interpretation:} Network dispatch issues such as feeder congestion and bus voltage limits could not directly specify operating instructions that a BEM can execute. The same operational intent therefore involves different objects, variables, units, time scales, and constraints across buildings and power grid domains. The challenge is to map the DNO's intent reliably into a task understandable to BEMs.
    \item \textbf{Cross-domain decision making:} Cross-domain decisions require coordinated use of both building and grid models. Conventional workflows involve manual model selection, input preparation, time-series alignment, building-to-bus mapping, and repeated execution. The challenge is to construct and execute a computational workflow from the task requirements using cross-field models and software.
\end{itemize}

\section{Methodology}

\subsection{Overall Workflow}

\begin{figure*}[t]
    \centering
    \includegraphics[width=1.7\columnwidth]{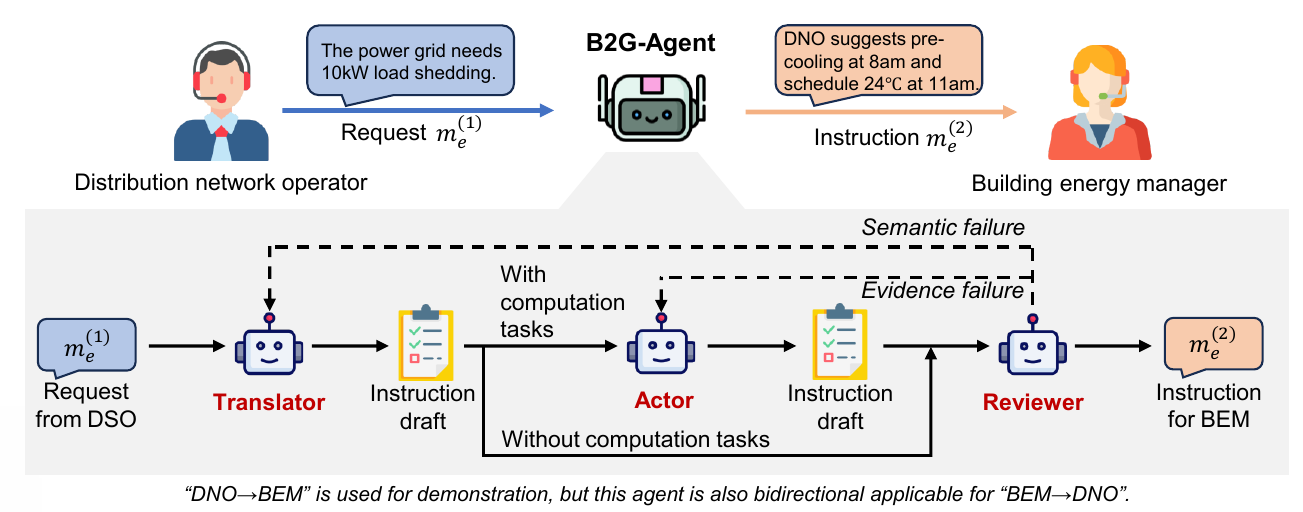}
    \caption{Overall architecture and end-to-end interaction workflow of the proposed B2G-Agent.}
    \label{framework}
    \vspace{-1.0em}
\end{figure*}

Fig.~\ref{framework} presents the proposed B2G-Agent for end-to-end building-grid interactions. It integrates three specialized agents as the core functions: the Translator, Actor, and Reviewer. For a complete interaction $e$, let $\ell\in\{1,2\}$ index the two cross-domain transfers: $\ell=1: m_e^{(1)}\!\rightarrow\!m_e^{(2)}$, and $\ell=2: m_e^{(3)}\!\rightarrow\!m_e^{(4)}$. The direction of each transfer is denoted by $\delta_{e,\ell}\in\{\mathrm{B2G},\mathrm{G2B}\}$, denoting building-to-grid (B2G) and grid-to-building (G2B) transfers, respectively. The information exchanged among agents in one single direction $\ell$ is represented by the set $\mathcal{R}_{e,\ell}$ as
\begin{equation}
\mathcal{R}_{e,\ell}=
\left(
\widetilde m_e^{(2\ell)},
\boldsymbol z_{e,\ell},
\mathcal Y_{e,\ell}^{\mathrm{sim}},
\mathcal L_{e,\ell}
\right),
\label{eq:internal-record}
\end{equation}
where $\widetilde m_e^{(2\ell)}$ is a natural-language instruction draft for the input message; $\boldsymbol z_{e,\ell}$ is the structured task state read and processed by agents; $\mathcal Y_{e,\ell}^{\mathrm{sim}}$ is the evidence set of simulation and optimization results; $\mathcal L_{e,\ell}$ records accompanying information for agent processing, including knowledge sources, tool calls, and execution status.

The Translator first reads the input message $m_e^{(2\ell-1)}$ and performs direction-dependent cross-domain semantic alignment to generate $\boldsymbol z_{e,\ell}$ and $\widetilde m_e^{(2\ell)}$. If the task state requires computation, the Actor converts it into an executable tool plan and invokes optimization and simulation software. It obtains $\mathcal Y_{e,\ell}^{\mathrm{sim}}$, records $\mathcal L_{e,\ell}$, and writes the results back to the task state and instruction draft. If no computation is required, the task proceeds directly to the Reviewer. The Reviewer jointly examines $m_e^{(2\ell-1)},\boldsymbol z_{e,\ell},\widetilde m_e^{(2\ell)},\mathcal Y_{e,\ell}^{\mathrm{sim}},\mathcal L_{e,\ell}$ to identify semantic or evidence failures. Only drafts that pass review are released as the external message $m_e^{(2\ell)}$.

More detailed functions for each agent are introduced in the following subsections.

\subsection{Translator Agent: Bidirectional Semantic Alignment}

\begin{figure}[t]
    \centering
    \includegraphics[width=0.8\columnwidth]{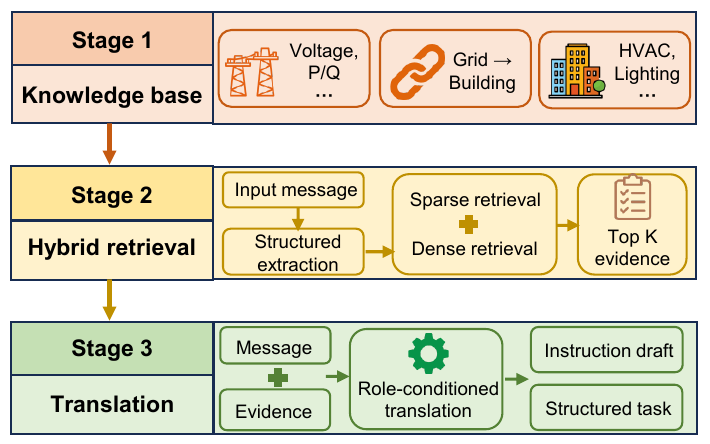}
    \caption{Key steps of knowledge-base construction, hybrid retrieval, and semantic translation in the Translator agent.}
    \label{translator}
    \vspace{-1.0em}
\end{figure}

\begin{figure}[t]
    \centering
    \includegraphics[width=1.0\columnwidth]{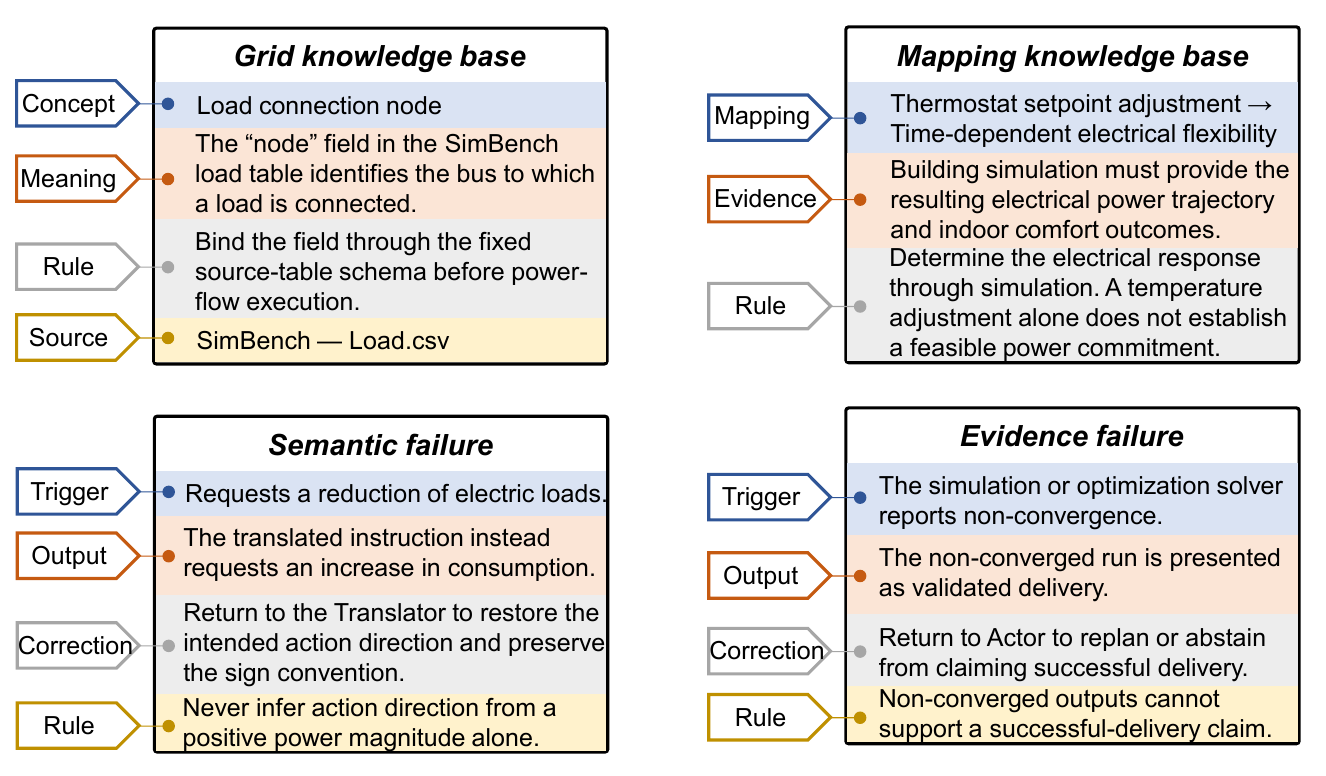}
    \caption{Structured records for domain knowledge, cross-domain mappings, and semantic and evidence failures.}
    \label{knowledge}
    \vspace{-1.0em}
\end{figure}

Fig.~\ref{translator} illustrates the Translator's major workflow. We first construct a dual-domain knowledge base $\mathcal D$ for building-grid interaction tasks, with examples shown in Fig.~\ref{knowledge}. Examples of building-domain knowledge include thermal comfort, heating, ventilation, and air conditioning (HVAC) precooling and preheating, equipment operating constraints, thermal inertia, and occupant behavior. Examples of grid-domain knowledge include active and reactive power, bus voltage constraints, adjustment directions, and scheduling time scales. Cross-domain mapping examples describe how building concepts translate into bus- and time-specific adjustment bounds and which building operations are relevant to grid objectives. Each knowledge record stores structured attributes, including name, definition, units, time scale, and spatial level.

For input message $m_e^{(2\ell-1)}$, the Translator first extracts structured attributes under the field schema $\mathcal S_{e,\ell}$ associated with direction $\ell$. It then applies sparse retrieval and embedding-based dense retrieval separately to source-domain knowledge, target-domain knowledge, and cross-domain mapping examples~\cite{robertson2009bm25,karpukhin2020dpr}. Sparse retrieval matches technical terms, variable names, and units, while dense retrieval identifies synonymous expressions and natural-language paraphrases. Let $d\in\mathcal D$ denote a knowledge record, 
the score $s_{e,\ell}(d)$ of candidate record $d$ is obtained by weighted reciprocal rank fusion (RRF)~\cite{cormack2009rrf}. Let $\mathcal D_{e,\ell}^{\mathrm{cand}}$ contain only records that appear in at least one retrieval list. The retrieved context $\mathcal K_{e,\ell}^{\mathrm{ret}}$ is then extracted as the top $K$ highest-scoring eligible records.


The Translator subsequently generates the following outputs through schema-constrained LLM generation:
\begin{equation}
\left(
\boldsymbol z_{e,\ell}^{(0)},
\widetilde m_e^{(2\ell)}
\right)
=
\mathcal T_{\theta}
\left(
m_e^{(2\ell-1)},
\mathcal K_{e,\ell}^{\mathrm{ret}};
\mathcal S_{e,\ell}
\right),
\label{eq:translator}
\end{equation}
where $\mathcal T_{\theta}$ denotes the Translator using an LLM with frozen parameters $\theta$, and $\boldsymbol z_{e,\ell}^{(0)}$ is the initial structured task state before Actor processing. Each field is assigned one of four status labels: \texttt{declared} identifies information provided directly by the sender; \texttt{mapped} identifies semantic or unit mappings supported by reviewed knowledge; \texttt{to-be-simulated} identifies fields $\mathcal F_{e,\ell}^{\mathrm{sim}}$ requiring Actor computation; \texttt{needs-clarification} identifies information $\mathcal F_{e,\ell}^{\mathrm{clar}}$ that should be obtained from the sender. After the required clarifications are resolved, if $\mathcal F_{e,\ell}^{\mathrm{sim}}$ is not empty, the internal record proceeds to the Actor for the following computation process.

\subsection{Actor Agent: Cross-Field Computation Orchestration}

\begin{figure}[t]
    \centering
    \includegraphics[width=1.0\columnwidth]{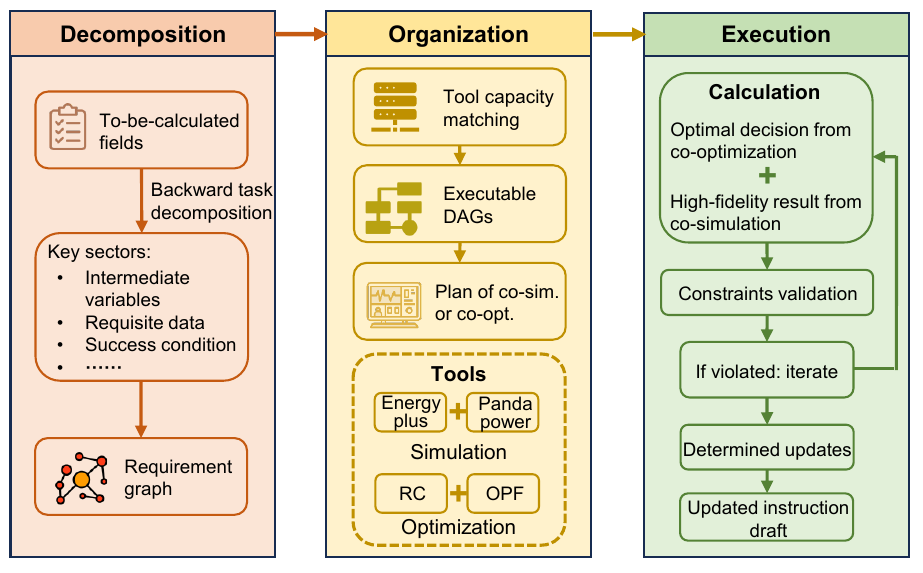}
    \caption{Key steps of instruction decomposition, and computation plan and execution in the Actor agent.}
    \label{actor}
     \vspace{-1.0em}
\end{figure}

Fig.~\ref{actor} illustrates the Actor's major workflow. The Actor maintains two building-grid model combinations with distinct functions. The EnergyPlus--pandapower tool-chain provides high-fidelity joint simulation~\cite{energyplusDocs,pandapowerDocs}. EnergyPlus simulates device schedules, energy use, indoor temperature, and thermal comfort in buildings, while pandapower performs power flow analysis to calculate variables such as bus voltages and network losses. Standardized time series and BEM-to-bus mapping connect the two simulators for sequential execution. In addition, the resistance-capacitance (RC)--optimal power flow (OPF) toolchain supports joint optimization. RC models represent building thermal dynamics and flexibility constraints at low computational cost. They can be embedded in OPF models with other flexible resources in buildings as linear constraints to generate candidate response schedules. The RC model parameters can be identified from EnergyPlus simulation data using the least-squares method~\cite{vinther2013rcIdentification}.

The Actor first reads all fields requiring computation in $\boldsymbol z_{e,\ell}^{(0)}$ and applies schema-guided backward task decomposition~\cite{zhou2023leastToMost,shen2023hugginggpt}. It traces each target field back to intermediate variables, prerequisite data, tool operations, and success criteria, yielding the computational requirement graph $\mathcal G_{e,\ell}^{\mathrm{req}}$. For example, determining whether a proposed adjustment causes a bus voltage violation requires buildings' power responses and nodal active and reactive loads. These inputs support power flow analysis to quantify bus voltages and line loading. To further generate a tool plan, the Actor matches $\mathcal G_{e,\ell}^{\mathrm{req}}$ against the typed tool capability graph $\mathcal G^{\mathrm{tool}}$. A tool registry stores the required structured attributes for each function. Let $\Omega(\mathcal G_{e,\ell}^{\mathrm{req}},\mathcal G^{\mathrm{tool}},\boldsymbol z_{e,\ell}^{(0)})$ denote the set of candidate tool execution plans represented as directed acyclic graphs (DAGs). Each candidate covers all fields requiring computation and satisfies tool-type compatibility, input availability, and data-dependency constraints. For a candidate plan $\Pi$, the node set $\mathcal V(\Pi)$ represents tool calls and the edge set $\mathcal E(\Pi)$ represents data dependencies from upstream outputs to downstream inputs or precedence constraints~\cite{kim2024llmCompiler}. The Actor selects the candidate plan $\Pi_{e,\ell}^{\star}$ as
\begin{equation}
\Pi_{e,\ell}^{\star}
=\arg\min_{\Pi\in\Omega(
\mathcal G_{e,\ell}^{\mathrm{req}},
\mathcal G^{\mathrm{tool}},
\boldsymbol z_{e,\ell}^{(0)})}
\sum_{u\in\mathcal V(\Pi)}\kappa_u,
\label{eq:tool-plan}
\end{equation}
where $u$ denotes a tool node and $\kappa_u\geq0$ is its normalized cost, including execution time and solution accuracy.

The resulting tool execution plan determines whether the task requires joint optimization, co-simulation, or sequential execution. An optimization output is a candidate response and must pass the required co-simulation checks before release. For joint optimization, $\Pi_{e,\ell}^{\star}$ specifies the participating BEMs, along with their RC and other models in building energy systems. 
Let $\boldsymbol u$ collect the control trajectories of all BEMs, and the joint optimization problem is represented as $\boldsymbol u_e^{\mathrm{opt}}
=\arg\min_{\boldsymbol u\in\mathcal U_e}(
J_e^{\mathrm{DNO}}(\boldsymbol u)
+\varpi J_e^{\mathrm{BEM}}(\boldsymbol u))$, where $\boldsymbol u_e^{\mathrm{opt}}$ is the candidate optimal response obtained by the joint optimization; $\mathcal U_e$ denotes the feasible set defined by the RC state equations with constraints (e.g., equipment capacities, flexibility bounds, storage energy constraints, and network constraints); $J_e^{\mathrm{DNO}}$ and $J_e^{\mathrm{BEM}}$ are scalar costs representing the operational objectives from both DNO and BEMs, respectively; $\varpi$ is the weight fixed during task generation to represent the trade-off between the two objective terms.

For the co-simulation, $\Pi_{e,\ell}^{\star}$ specifies baseline and response scenarios, the EnergyPlus models for each BEM, weather files, and control schedules. It also specifies simulation periods, output variables, temporal resampling and unit conversion rules, and the BEM-to-bus mapping. During co-simulation, the Actor applies each candidate control schedule to the corresponding EnergyPlus building models to obtain aggregated loads from buildings. At each time step, pandapower performs AC power flow analysis. It then uses these results, together with the EnergyPlus outputs, to verify security constraints in the distribution network.

After tool execution, deterministic routines collect unit conversion, temporal alignment, BEM-to-bus mapping, and field updates into an evidence package. The Actor marks successfully computed fields as \texttt{resolved}. Fields affected by software failures or insufficient inputs remain \texttt{unresolved}. The instruction draft $\widetilde m_e^{(2\ell)}$ is updated accordingly. The updated task state, instruction draft, and evidence package are then passed to the Reviewer.

\subsection{Reviewer Agent: Reflective Instruction Validation}

\begin{figure}[t]
    \centering
    \includegraphics[width=0.8\columnwidth]{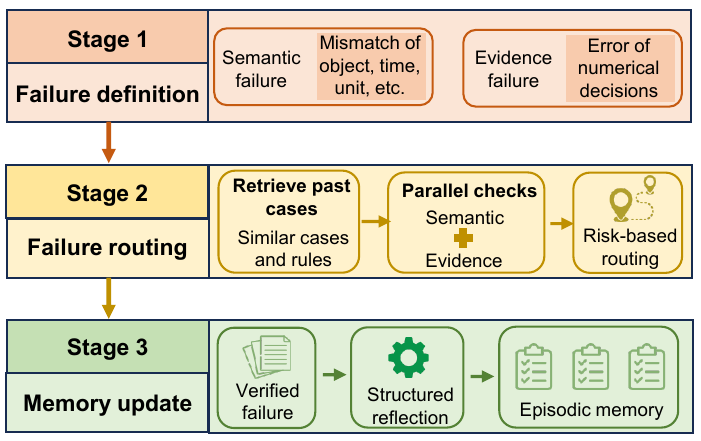}
    \caption{Key steps of failure definition and routing, and memory update in the Reviewer agent.}
    \label{reviewer}
     \vspace{-1.0em}
\end{figure}

Fig.~\ref{reviewer} illustrates the Reviewer's major workflow. Failures in interactions are classified as semantic or evidence failures. Semantic failures originate in the Translator and include omitted or reversed objectives, constraints, as well as incorrect BEM, power grid, or equipment identifications. Evidence failures originate in the Actor and include incorrect tool selection, execution order, or model-to-parameter bindings. Fig.~\ref {knowledge} shows examples of both failure types.

The Reviewer first retrieves similar historical failure cases $\mathcal M_{e,\ell}^{\mathrm{ret}}$ and their reflection rules from memory $\mathcal M^{\mathrm{mem}}$ based on the application, interaction direction, and current fields. The semantic channel uses a schema-constrained LLM critic to compare the original input, task state, and natural-language draft. The result of semantic check $j$ is defined as
\begin{equation}
\eta_{e,\ell,j}^{\mathrm{sem}}
=\phi_j
\left(
m_e^{(2\ell-1)},
\boldsymbol z_{e,\ell},
\widetilde m_e^{(2\ell)},
\mathcal M_{e,\ell}^{\mathrm{ret}}
\right)
\in\left\{0,\tfrac{1}{2},1\right\},
\label{eq:semantic-verdict}
\end{equation}
where $\phi_j(\cdot)$ is the schema-constrained LLM function for semantic check $j$; the values 0, $1/2$, and 1 indicate pass, uncertain, and fail, respectively.

The evidence channel uses deterministic validators to check schema validity, units, execution status, numerical results, and hard constraints. The result of evidence check $k$ is
\begin{equation}
\eta_{e,\ell,k}^{\mathrm{evi}}
=\psi_k
\left(
\boldsymbol z_{e,\ell},
\mathcal Y_{e,\ell}^{\mathrm{sim}},
\mathcal L_{e,\ell}
\right)
\in\left\{0,\tfrac{1}{2},1\right\},
\label{eq:evidence-verdict}
\end{equation}
where $\psi_k(\cdot)$ denotes the deterministic validation function for evidence check $k$. The three values have the same interpretation as in the semantic channel, where an uncertain result indicates that the available evidence is insufficient for a definitive check.

For channel $c\in\{\mathrm{sem},\mathrm{evi}\}$, the Reviewer aggregates the check results into a risk score:
\begin{equation}
\varrho_{e,\ell}^{c}
=\frac{
\sum_{j\in\mathcal J_c}
\omega_j^{c}\eta_{e,\ell,j}^{c}
}{
\sum_{j\in\mathcal J_c}\omega_j^{c}
},
\label{eq:risk-score}
\end{equation}
where $\mathcal J_c$ is the nonempty set of applicable checks in channel $c$, and $\omega_j^{c}\geq0$ is the weight of check $j$. The risk score $\varrho_{e,\ell}^{c}\in[0,1]$ summarizes failed and uncertain checks. A failed semantic check, an unresolved mandatory semantic field, or $\varrho_{e,\ell}^{\mathrm{sem}}$ higher than a preset threshold will return the task to the Translator. If both semantic and evidence failures occur, the Translator will first correct the task state, after which the Actor will recompute any affected results. A failed evidence check, an unresolved mandatory evidence field, an uncertain hard-constraint check, or $\varrho_{e,\ell}^{\mathrm{evi}}$ higher than a preset threshold will return the task to the Actor.

For confirmed failures, the Reviewer applies schema-constrained Reflexion~\cite{shinn2023reflexion} to generate structured reflection records. Each record contains structured failure traces, including trigger conditions, failure type, responsible agent, incorrect output, verified correction, generalizable rule, and applicability conditions. These records are added to $\mathcal M^{\mathrm{mem}}$ to update the historical failure-case repository during interactions.

\subsection{User Interface}

The developed tool provides a unified, role-conditioned user interface for practical applications of both DNOs and BEMs. Users first select their roles and submit operational requests in natural language. According to the direction, the interface displays the fields relevant to the sender's domain, while the recipient receives human-readable operating instructions. Each recipient receives information expressed in their own domain terminology and can confirm, correct, or reject it while focusing on their responsibilities.

\begin{figure*}[t]
    \centering
    \includegraphics[width=1.8\columnwidth]{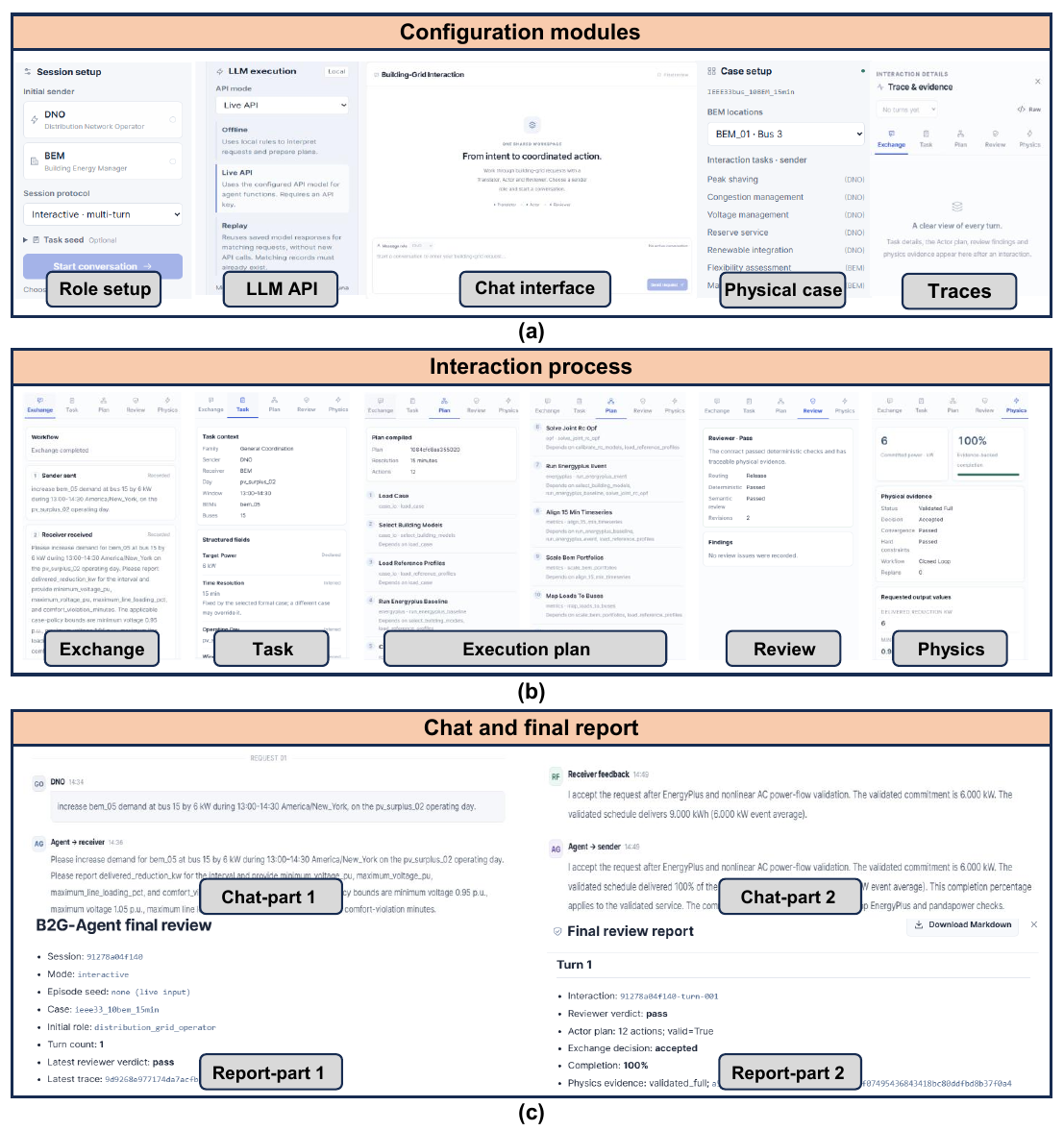}
    \caption{Interface including configuration modules, interaction process, and chat and final report of B2G-Agent.}
    \label{interface}
     \vspace{-1.0em}
\end{figure*}

Fig.~\ref{interface} presents the configuration modules, the interaction process, and the chat and final report information within the user interface. In particular, Fig.~\ref{interface}a shows the role setup, LLP settings, chat workspace, physical case, and trace panel. Users can select the role (DNO or BEM), the interaction protocol, and the LLM execution mode. The case panel identifies the selected network and the locations of BEMs, and the trace panel provides access to intermediate records for each turn of tasks. Fig.~\ref{interface}b shows five trace views: natural-language messages, structured task fields, Actor tool calls, Reviewer routing decisions, and simulation or optimization outputs. Fig.~\ref{interface}c illustrates the detailed chat information and the final report with the case configuration, interaction history, final decision, and physical evidence.

\section{Case Studies}\label{sec_Case}

\subsection{Experimental Setups}

The case studies use the IEEE 33-Bus Test System as the distribution network with 10 BEMs connected to load buses. Grid inputs include network topology, line impedances and current ratings, and 15~min nodal active and reactive load profiles. Additional inputs include photovoltaic (PV) forecasts, storage states of charge, and charging/discharging limits. We obtain bus voltages, branch power flow, and network losses from power flow analysis in pandapower~\cite{pandapowerDocs}. Each BEM is represented by 1--5 EnergyPlus building prototypes to characterize the equivalent response of a large building or a building cluster. Model inputs include EnergyPlus input data files, weather files, occupancy and equipment schedules, thermal comfort ranges, controllable equipment, and baseline and response control strategies~\cite{energyplusDocs}. Each interaction is evaluated over a day with a uniform 15~min time resolution. The scenarios cover load peaks, PV surplus periods, local voltage violations, overload risks, and multiple candidate response windows. These conditions provide the physical setting for the interaction tasks generated in the following subsection.

All three agents in the tool use the same OpenAI \texttt{gpt-6-luna} model. Key parameters such as temperature and maximum tokens per task are all fixed in case studies. Table~\ref{tab:framework_configuration} summarizes these key parameters in the workflow.
\begin{table}[t]
    \centering
    \caption{Critical Parameters of the Proposed Tool}
    \label{tab:framework_configuration}
    \resizebox{1\columnwidth}{!}{
    \begin{tabular}{lccc}
        \toprule
        \textbf{Parameter} & \textbf{Value} & \textbf{Description} \\
        \midrule
        Temperature
        & 0
        & Randomness \\

        Top-p
        & 1.0
        & Retains all candidate tokens \\

        Maximum Tokens
        & 9000
        & Tokens for LLM outputs per call \\

        Embedding model
        & \texttt{all-MiniLM-L6-v2}
        & Semantic knowledge retrieval \\

        Embedding dimension
        & 384
        & Number of vector components \\
        \bottomrule
    \end{tabular}}
    \vspace{-1.0em}
\end{table}

All physical simulations run on a laptop with a 13th-generation Intel processor at 2.60~GHz and 16~GB of RAM. Source code and case study results are available on GitHub~\cite{b2g-agent} and also on the PowerAgent community~\cite{poweragent}.

\subsection{Task Generation and Evaluation}

\subsubsection{Task Generation}

Each task is defined as a bidirectional exchange comprising messages $m_e^{(1)}$ through $m_e^{(4)}$. Dispatch signals and price signals from demand response serve as the primary interaction mechanisms. Seven task categories are considered: \textit{Peak shaving} tasks evaluate reductions in feeder peak demand; \textit{Congestion management} tasks evaluate the restoration of overloaded lines to acceptable loading levels; \textit{Voltage management} tasks evaluate the elimination of bus voltage violations; \textit{Reserve service} tasks consider capacity declaration, reservation, and event activation at a 15~min resolution; \textit{Renewable integration} tasks evaluate reductions in PV curtailment; \textit{Flexibility assessment} tasks evaluate baselines and adjustment bounds of building loads; \textit{Market trading} tasks follow a simplified process of nodal flexibility bids with prices and settlement rules. BEMs typically initiate baseline reports, flexibility declarations, and offers, whereas the DNO typically initiates dispatch tasks. We generate a total of 200 interaction tasks for the study. These comprise 150 single-application tasks across seven categories and 50 composite tasks involving multiple categories. Various tasks involve single or multiple BEMs, single or multiple time intervals, and both feasible and infeasible requests. The task list is available at \path{tasks/episodes.jsonl} in the GitHub repository.

Along with interaction tasks, we construct the dual-domain knowledge base required by Translator and the initial failure-case repository required by Reviewer for the study. The knowledge base contains building-domain resources, grid-domain resources, and cross-domain mapping examples. The knowledge base contains building-domain resources, grid-domain resources, and cross-domain mapping examples. We extract the domain resources from frozen public datasets and software repositories. Cross-domain mappings combine source-derived examples with predefined conversion rules. To generate the initial failure-case repository, we introduce controlled perturbations corresponding to semantic and evidence failures into tasks, then execute the full workflow. Only failures independently confirmed by validation programs and accompanied by verified corrections are stored in memory. The knowledge records and initial failure cases are stored at \path{knowledge/} and \path{verified_seed.jsonl}, respectively. Note that online memory is disabled during benchmark evaluation, so the Reviewer retrieves only from the same read-only seed repository across tasks and repetitions.

\subsubsection{Evaluation Metrics}

The evaluation covers three aspects: interaction quality, agent performance, and LLM usage.

The quality of each task $e$ is scored by four components ranging from $[0,1]$: bidirectional semantic fidelity $Q^F_e$, interaction achievement $Q^A_e$, physical feasibility $Q^V_e$, and objective completeness $Q^G_e$. All metrics show better performance with higher values.

Semantic fidelity evaluates both cross-domain transfers $\ell=1,2$. Let $\mathcal F_{e,\ell}^{\mathrm{sem},\star}$ and $\widehat{\mathcal F}_{e,\ell}^{\mathrm{sem}}$ denote the reference and final output sets of semantic field identifiers, respectively. We first quantify the weighted semantic precision $F_{e,\ell}^\text{pre}$ and recall $F_{e,\ell}^\text{rec}$ rates by comparing reference and final output sets, and $F_e$ is
\begin{equation}
Q^F_e=\sum_{\ell=1}^{2}{
F_{e,\ell}^\text{pre}F_{e,\ell}^\text{rec}
/(F_{e,\ell}^\text{pre}+F_{e,\ell}^\text{rec})}.
\end{equation}

Interaction achievement evaluates whether the recipient can act on the instruction and whether the sender can confirm the outcome after interactions. Let $\mathcal F_e^{\mathrm{req}}$ denote the set of required field instances across the two output messages. Applicable fields include the target bus or BEM, adjustment direction, quantity, operating constraints, and status. The indicator $\iota_{e,j}^{\mathrm{req}}=1$ if field $j$ is unambiguous and compatible. Then the score $Q^A_e$ is defined as
\begin{equation}
Q^A_e=
\sum_{j\in\mathcal F_e^{\mathrm{req}}}
\beta_j^{\mathrm{req}}\iota_{e,j}^{\mathrm{req}}
/\sum_{j\in\mathcal F_e^{\mathrm{req}}}
\beta_j^{\mathrm{req}},
\end{equation}
where $\beta_j^{\mathrm{req}}>0$ is the weight of required field $j$.

Physical feasibility is evaluated by hard-constraint violations. Let $\mathcal C_e^{\mathrm{hard}}$ denote the set of hard constraints from both the building and power grid sides, expressed as $g_{e,h}(\cdot)\leq0$ with index $h$. The score $Q^V_e$ is defined as
\begin{equation}
Q^V_e=\prod_{h\in\mathcal C_e^{\mathrm{hard}}}\mathbb I\!\left[
g_{e,h}(\widehat{\boldsymbol u}_e)\leq\epsilon_h
\right],    
\end{equation}
where $\epsilon_h$ is a small-value tolerance, $\widehat{\boldsymbol u}_e$ is the final decisions after interactions, and $\mathbb I(\cdot)$ equals 1 when its condition holds and 0 otherwise.

Objective completeness measures improvement of interaction outputs relative to the reference. Let $\boldsymbol y$ denote the task outcome: a control trajectory for dispatch tasks or a numerical estimate for assessment tasks. We denote the proposed, baseline, and ideal outcomes by $\widehat{\boldsymbol y}_e$, $\boldsymbol y_e^0$, and $\boldsymbol y_e^{\star}$, respectively. Let $o\in\mathcal O_e$ index the set of operational objectives, and let $D_{e,o}(\boldsymbol y)\geq0$ denote the residual shortfall for outcome $\boldsymbol y$. For peak shaving, $D_{e,o}$ is the unmet peak-demand reduction. For congestion and voltage management, it is the remaining time-integrated constraint violation. For reserve services, it is a weighted sum of shortfalls in capacity and duration. For renewable energy integration, it is the remaining curtailed PV energy. For flexibility assessment, it is the normalized estimation error. For market trading, it is the shortfall in delivered quantity relative to the contract. The score $Q^G_e$ is defined as
\begin{equation}
    Q^G_{e}=\sum_{o\in\mathcal{O}_e}\operatorname{clip}
\left(
\frac{
D_{e,o}(\boldsymbol y_e^0)
-D_{e,o}(\widehat{\boldsymbol y}_e)
}{
D_{e,o}(\boldsymbol y_e^0)
-D_{e,o}(\boldsymbol y_e^{\star})
},0,1
\right),
\end{equation}
where $\operatorname{clip}(x,0,1)=\min\{1,\max\{0,x\}\}$ to bound $x$ within $[0,1]$.

Combining the above components of quality, the total quality score on a [0,1] scale is defined as
\begin{equation}
Q^\text{qua}_e=w^FQ^F_e+w^AQ^A_e+w^VQ^V_e+w^GQ^G_e
\label{quality}
\end{equation}
with $\sum_{i=1}^{4}{w^i}=1, \forall w^i>0, i\in\{F,A,V,G\}$.

The task success indicator is further defined as
\begin{equation}
Q^\text{tsr}_e
=\mathbb I(Q^F_e\geq\tau_F)
\mathbb I(Q^A_e=1)
Q^V_e
\mathbb I(Q^G_e\geq\tau_G),
\label{success}
\end{equation}
where $\tau_F,\tau_G$ are the semantic fidelity and objective completeness thresholds, respectively.

Let $N^{\mathrm{epi}}$ denote the number of evaluated tasks ($N^{\mathrm{epi}}=200$ as we introduced above), and each task is repeated $N^{\mathrm{rep}}$ times ($N^{\mathrm{rep}}=3$ in our study for result robustness). The mean interaction quality $Q^\text{qua}$ and task success rate (TSR) $Q^\text{tsr}$ are calculated by
\begin{equation}
Q^\text{qua/tsr}
=\frac{1}{N^{\mathrm{epi}}N^{\mathrm{rep}}}
\sum_{e=1}^{N^{\mathrm{epi}}}
\sum_{r=1}^{N^{\mathrm{rep}}}
Q^\text{qua/tsr}_{e,r}
\end{equation}
with $e,r$ as indices for tasks and repetitions, respectively.

Furthermore, for the evaluation of agents' functional performance, three metrics are reported for the Translator on its first forward draft: weighted semantic-field F-score $P^\text{tra}_1$; cross-domain mapping accuracy $P^\text{tra}_2$, jointly checking objects, variables, directions, units, and time scales; and clarification-decision F-score $P^\text{tra}_3$. For the Actor's terminal candidate after revisions, plan validity rate $P^\text{act}_1$ checks DAG and tool/input compatibility; tool-chain success rate $P^\text{act}_2$ requires completed calls and requested outputs; and evidence consistency rate $P^\text{act}_3$ checks numerical tolerances and hard-constraint verdicts. Reviewer feedback is evaluated by the appropriate first-review handling rate $P^\text{rew}_1$, requiring successful clarification, clean progression to reverse review, or verified draft correction; the reverse-review reach rate $P^\text{rew}_2$ over the same source-executable runs; and task success rate $P^\text{rew}_3$ over the same source-executable runs. All metrics indicate that higher values are better performance.

Finally, we evaluate LLM usage performance by total token consumption per task, LLM inference latency, and retrieval time for the overall process.

\subsubsection{Comparisons}
\begin{itemize}
    \item \textit{Ground-truth:} The ground-truth results provide references for interaction tasks, which are generated through a manually specified and independently validated optimization or simulation procedure. The ground-truth results are stored in the \path{groundtruth/} directory of the repository.
    \item \textit{Baseline:} The baseline method is a computational tool-enabled single agent that uses the same LLM. It has access to the same dual-domain knowledge base and software tools. However, it omits the Translator--Actor--Reviewer role decomposition and reflective online memory. A single agent performs interpretation, tool invocation, and output generation.
    \item \textit{Ablations:} We use ablation methods to evaluate the three agents separately. For Translator, plain translation removes retrieval, semantic prefilling, and object pairing. For Actor, unstructured planning removes task-specific DAGs, parameter prefilling, schema-constrained generation, and automatic normalization. For Reviewer, no-feedback disables corrective returns to Translator/Actor while retaining review, memory, and release gates.
\end{itemize}

The evaluations for all comparison methods are listed in Table~\ref{tab-comp}, where metrics for interaction quality are $\{Q^\text{qua},Q^\text{tsr},Q^\text{F},Q^\text{A},Q^\text{V},Q^\text{G}\}$, and metrics for agents are $\{F^\text{tra}_{1,2,3},F^\text{act}_{1,2,3},F^\text{rew}_{1,2,3}\}$.
\begin{table}[t]
\caption{Evaluation of Comparison Methods}
\centering
\resizebox{1.0\linewidth}{!}{
\begin{tabular}{l|ccccc}
\hline
\multicolumn{1}{c|}{\multirow{2}{*}{}} & \multicolumn{1}{c}{\multirow{2}{*}{Interaction quality}} & \multicolumn{3}{c}{Agent performanc}                                                                                         & \multicolumn{1}{c}{\multirow{2}{*}{LLM usage}} \\ \cline{3-5}
\multicolumn{1}{c|}{}                  & \multicolumn{1}{c}{}                                     & \multicolumn{1}{c}{Translator} & \multicolumn{1}{c}{Actor} & \multicolumn{1}{c}{Reviewer} & \multicolumn{1}{c}{}                                  \\ \hline
Proposed                               &  \checkmark                                                        &     \checkmark                                       &          \checkmark                             &                       \checkmark                   &  \checkmark                                                     \\
Baseline                               &    \checkmark                                                      &  $\times$                                          &                    $\times$                   &    $\times$                                      &    \checkmark                                                   \\
Ablation-Translator                    &     \checkmark                                                     &          \checkmark                                  &                          $\times$             & $\times$                                         &         \checkmark                                              \\
Ablation-Actor                         &          \checkmark                                                &       $\times$                                     &                  \checkmark                     &  $\times$                                        &       \checkmark                                                \\
Ablation-Reviewer                      &        \checkmark                                                  &     $\times$                                       &         $\times$                              &        \checkmark                                  &          \checkmark                                             \\ \hline
\end{tabular}}
\label{tab-comp}
\vspace{-1.0em}
\end{table}

\subsection{Results of Building-Grid Interactions}

Table~\ref{bem-exchange} presents a complete interaction of congestion management within the preset tasks. This example illustrates the end-to-end interaction workflow enabled by the proposed tool. We use the interaction outputs to evaluate the interaction quality, agent performance, and LLM usage as follows.

\begin{table}[t]
\caption{A Typical Interaction Task of Congestion Management}
\label{bem-exchange}
\centering
\footnotesize
\renewcommand{\arraystretch}{1.12}
\begin{tabular}{@{}p{0.12\columnwidth}@{}%
                >{\raggedright\arraybackslash}p{0.88\columnwidth}@{}}
\toprule
$m_e^{(1)}$ &
\textbf{DNO request.} Ask BEM~07 at bus~21 to reduce demand by
210~kW from 16:00 to 18:00. \\

$m_e^{(2)}$ &
\textbf{Instruction to the BEM.} Assess the requested reduction
against the voltage, line-loading, and comfort limits. \\

$m_e^{(3)}$ &
\textbf{BEM response.} Offer 47.9~kW of sustained reduction
against the 210~kW request, with a window-average reduction
of 189.7~kW. \\

$m_e^{(4)}$ &
\textbf{DNO verification.} The schedule provided
379.5~kWh of simulated load reduction (90.35\% of the requested energy).
During the window, voltages stayed within 0.983--1.009~p.u.,
and peak line loading fell from 41.85\% to 40.78\% with no overloads.
EnergyPlus showed 0~min of response-induced comfort violations. \\
\bottomrule
\end{tabular}
\vspace{-1.0em}
\end{table}

\subsubsection{Quality Scores of Interaction Tasks}
\begin{figure}[t]
    \centering
    \includegraphics[width=1.0\columnwidth]{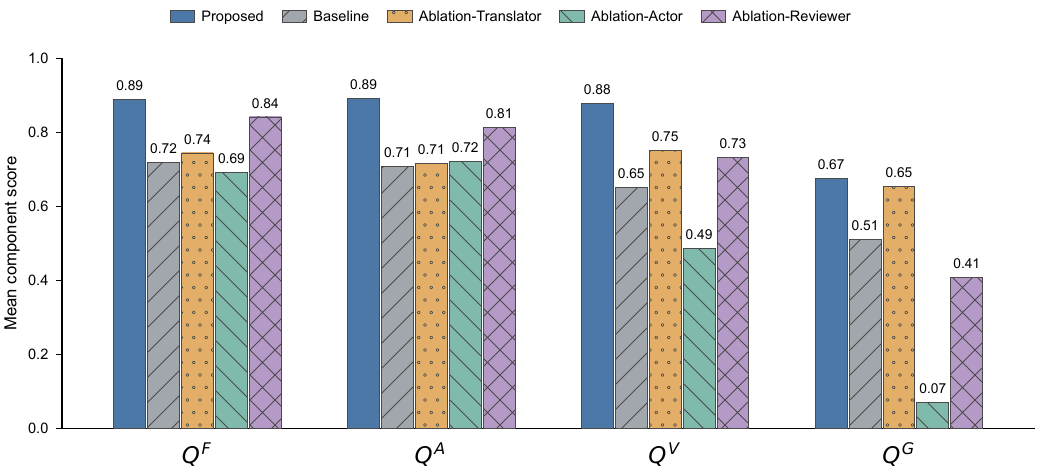}
    \caption{Performance of quality components across comparison methods.}
    \label{per-component}
    \vspace{-1.0em}
\end{figure}
Fig.~\ref{per-component} shows that the proposed tool achieves the highest mean score for each quality component. The Translator ablation brings semantic fidelity ($Q^F$) and interaction achievement ($Q^A$) close to the single-agent baseline. The Actor ablation produces the largest losses in physical feasibility ($Q^V$) and objective completeness ($Q^G$), falling from 0.88 to 0.49 and from 0.67 to 0.07, respectively. This pattern is consistent with removing the structured planning and input preparation needed to execute compatible tools and obtain valid cross-domain results. The Reviewer ablation retains diagnosis and release checks but disables corrective routing. It has lower scores than the proposed tool because detected translation or evidence failures cannot trigger revision by the corresponding agent.

\begin{figure}[t]
    \centering
    \includegraphics[width=1.0\columnwidth]{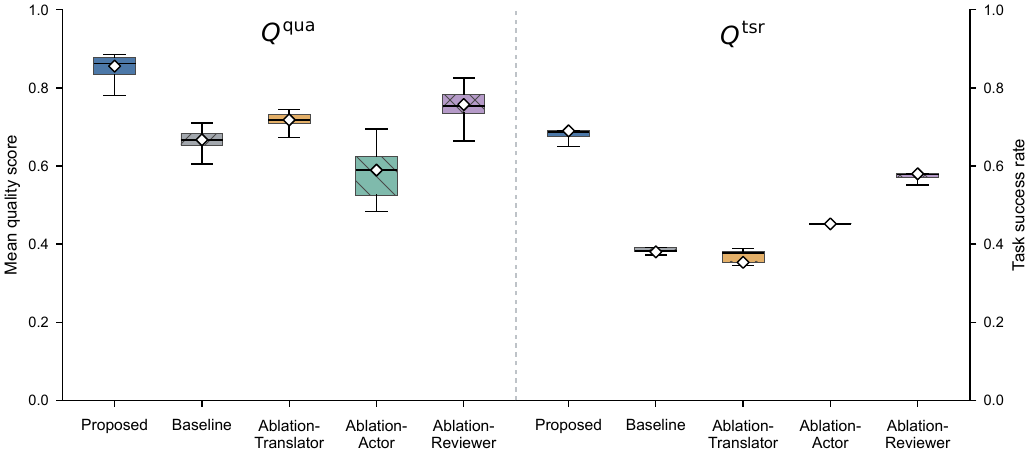}
    \caption{Sensitivity performance of task qualities across comparison methods given different combinations of weights in~\eqref{quality} and thresholds in~\eqref{success}.}
    \label{per-quality}
    \vspace{-1.0em}
\end{figure}
As shown in Fig.~\ref{per-quality}, the proposed tool achieves the highest performance across all tested weight and threshold settings. Its mean quality score ranges from 0.78 to 0.89, whereas the baseline ranges from 0.61 to 0.71. The Actor ablation shows the largest variation in quality as the weights change. This is consistent with its uneven component scores, particularly its low physical feasibility and objective completeness. For task success, the proposed tool maintains a rate of 0.65--0.69, compared with 0.37--0.39 for the baseline. The ordering of all five methods remains unchanged across the tested threshold settings. In particular, the Translator ablation has a lower success rate than the baseline despite its higher mean quality. While the weighted quality score allows stronger components to offset weaker ones, task success requires every applicable condition to be satisfied. Better physical results can therefore improve the quality score without resolving an incomplete interaction or insufficient semantic fidelity.

\subsubsection{Functional Performance of Agents}
\begin{figure*}[t]
    \centering
    \includegraphics[width=1.8\columnwidth]{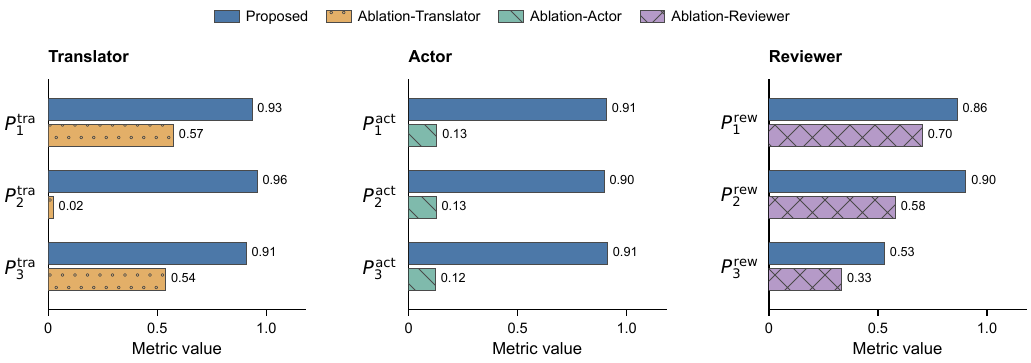}
    \caption{Comparison of agent functional performance between ablations and the proposed tool.}
    \label{per-agent}
     \vspace{-1.0em}
\end{figure*}
Fig.~\ref{per-agent} compares the proposed tool with the corresponding ablation for each agent. Without retrieval, semantic prefilling, and object pairing, the Translator's semantic-field and clarification-decision F-scores ($P^\text{tra}_1, P^\text{tra}_3$) decrease to 0.57 and 0.54, respectively. Cross-domain mapping $P^\text{tra}_2$ drops from 0.96 to 0.02. This metric requires the mapped objects, variables, directions, units, and time scales to agree jointly, making it sensitive to errors that a field-level score can tolerate.

The Actor ablation reduces plan validity, tool-chain success, and evidence consistency from approximately 0.90--0.91 to 0.12--0.13. Without task-specific execution graphs, schema constraints, parameter prefilling, and input normalization, the plan must infer the tool dependencies and required inputs directly from the task. The simultaneous declines in all three metrics show that the degradation extends from plan construction to execution and evidence generation.

The Reviewer ablation yields scores of $P^\text{rew}_1=0.70$ for appropriate first-review handling, $P^\text{rew}_2=0.58$ for reaching reverse review, and $P^\text{rew}_3=0.33$ for task success on source-executable runs. The corresponding scores are 0.86, 0.90, and 0.53 in the proposed tool. Review and memory remain active in this ablation, but the workflow cannot return a semantic or evidence failure to the Translator or Actor for correction. Consequently, an unresolved issue in the forward exchange can prevent a run from progressing to reverse review.

\subsubsection{Performance of LLM Usage}

Table~\ref{tab:computational-cost} reports mean resource use per task. The baseline uses fewer tokens than the proposed tool (65968.16 vs. 85214.18), but has higher LLM inference latency (101.12s vs. 80.15s) and retrieval time (0.93s vs. 0.72s). Both configurations use the same LLM, so these differences characterize their respective workflows. The Translator ablation reduces retrieval time to 0.03~s after removing knowledge retrieval, but uses around 2.4 times as many tokens and incurs around 2.1 times the LLM latency of the full workflow. The Actor ablation exhibits the same direction of change with smaller differences. These patterns are consistent with shifting work from retrieval and structured preparation to LLM inference, including any subsequent corrections. The Reviewer ablation reduces token consumption, inference latency, and retrieval time jointly. Disabling corrective returns eliminates the associated reprocessing, but reduces the quality and success rates in Fig.~\ref{per-quality}.
\begin{table}[t]
\centering
\caption{Performance of LLM Usage Across Comparison Methods}
\label{tab:computational-cost}
\resizebox{\columnwidth}{!}{%
\begin{tabular}{lrrr}
\toprule
Method
& \makecell{Tokens\\(per task)}
& \makecell{LLM latency\\(s/task)}
& \makecell{Retrieval\\(s/task)} \\
\midrule
Proposed            & 85214.18  & 80.15  & 0.72 \\
Baseline            & 65968.16  & 101.12 & 0.93 \\
Ablation-Translator & 208138.77 & 171.12 & 0.03 \\
Ablation-Actor      & 100040.69 & 100.80 & 0.31 \\
Ablation-Reviewer   & 35064.33  & 52.78  & 0.21 \\
\bottomrule
\end{tabular}%
}
\vspace{-1.0em}
\end{table}

\section{Conclusions}

This work presented B2G-Agent as an LLM-based multi-agent tool for end-to-end building-grid interactions. The Translator performs bidirectional semantic alignment through role-conditioned dual-domain hybrid RAG. The Actor uses schema-constrained task decomposition to plan and execute joint optimization or co-simulation. The Reviewer combines semantic and evidence checks with reflective memory to identify failures and route them for correction. Case studies with 200 synthesized interaction tasks evaluate interaction quality, agent performance, and LLM usage, showing that the proposed tool achieves a mean quality score of 85.57 and a task success rate of 0.69, higher than baseline and ablation methods. The closed-loop workflow across three specialized agents abstracts away cross-field modeling complexities and enables both DNOs and BEMs to focus on high-level operational decisions.

Future work will investigate the effectiveness under practical validations on the basis of simulated tasks, and will develop explanation functions tailored to decisions that are made for both DNOs and BEMs.

\section*{Acknowledgment}
We thank the PowerAgent community for their support in workflow development and suggestions.

\bibliographystyle{IEEEtran}
\bibliography{Ref}
\end{document}